\documentclass[a4paper,fleqn,usenatbib]{mnras} 

\pdfoutput=1
\usepackage[T1]{fontenc}
\usepackage{ae,aecompl}
\usepackage[export]{adjustbox}
\usepackage{graphicx}
\usepackage{times}

\usepackage[pscoord]{eso-pic}
\usepackage{amsmath}
\usepackage{amssymb}

\def\gsim{ \lower .75ex \hbox{$\sim$} \llap{\raise .27ex \hbox{$>$}} }
\def\lsim{ \lower .75ex \hbox{$\sim$} \llap{\raise .27ex \hbox{$<$}} }

 \title[Lava as warm dark matter]{Could fresh lava be (warm) dark matter?}
\author[M.~R.~Lovell et al.]{Mark R. Lovell\thanks{email: lovell@hi.is}$^{1}$\\
$^{1}$Center for Astrophysics and Cosmology, Science Institute, University of Iceland, Dunhagi 5, 107 Reykjavik, Iceland}

\date{Accepted ... Received ...; in original form ...} 
\pubyear{2022}

\begin{document}

\label{firstpage}
\pagerange{\pageref{firstpage}--\pageref{lastpage}} 
  
\maketitle

\begin{abstract}

\noindent Dark matter models can be classified according to their impact on the properties of galaxies, including cold dark matter (CDM), warm dark matter (WDM), self-interacting dark matter (SIDM) and fuzzy dark matter (FDM). In celebration of April Fool's Day, and also of the 1-year anniversary of the start of the 2022 volcanic eruption at Fagradalsfjall here in Iceland, we explore fresh lava as a candidate for WDM specifically. We verify first hand that lava is indeed warm (exhibits free-streaming and retains temperature for several months after the eruption ends, is 1000~K, sets fire to grass, one can feel one's eyebrows singe at a distance of 4~m) and dark once sufficiently decoupled from its source of production.   

\end{abstract}


\section{Introduction}
\label{intro}

The identity of the dark matter remains one of the most pressing problems in physics. Among its primary characteristics are that it be massive and that it not emit light; we also desire that it be detectable in experiments. The weakly interacting massive particle (WIMP) candidate is ripe for detection through its weak interactions with standard model particles. Direct detection experiments are typically located deep underground  in order to filter out noise from cosmic rays that mimic WIMP detection signals. Such experiments are expensive and inspire a degree of anxiety in those of us who harbour concerns of being stuck underground, while also having struggled to make a definitive detection of dark matter. 

We are therefore left considering alternative options to this approach, starting with one simple suggestion: if we cannot find the dark matter by going underground, why not instead  allow dark matter to come from deep underground up to us? An exciting opportunity of this kind presented itself on the 19th of March 2021, when a volcanic eruption began at Fagradalsfjall, 35~km from Reykjav\'ik. Magma originating at a depth of some 20~km below the Earth's surface -- compared to only 1.49~km for the Sanford laboratory where the LZ direct-detection experiment is situated -- promptly burst out of the ground, lighting up the night sky and pumping out unpleasant gases before cooling into dark rock. Given that this new hours old-rock is massive and does not emit light post cooling, we can envisage this fresh lava as a dark matter candidate. The high temperatures of this candidate also suggested that it could be considered specifically to be warm dark matter (WDM).

In order to investigate the value of this candidate as dark matter, we undertook two field trips in 2021 to gather evidence plus a series of observations over the following months up until the eruption ended in September 2021. This paper is organised as follows: we present the theory of lava as WDM, present the results of our field operations plus supplementary observations, and then draw appropriate conclusions.

\section{Theory of lava WDM} 
\label{rev}

The basic principle of dark matter is that it starts to cool immediately after production as it interacts with its surroundings. In the CDM-WIMP case, the dark matter exchanges energy with the primordial plasma through the weak nuclear force, and given the large mass of the dark matter particles relative to other standard model particles it cools particularly quickly and can form small dark matter haloes. In cases where the dark matter is light and not able to exchange energy with the primordial plasma, it can retain its large velocities for longer and so free stream out of small perturbations for longer: this is one definition of WDM.

Lava WDM achieves this separation from the plasma through inverse self-shielding: the cooling lava on the outside of the flow insulates the inner flow and thus prevents this inner flow from solidifying, thus enabling the lava to overcome the pull of gravity and flow out of small perturbations. Ultimately production will cease and the lava will cool sufficiently to solidify. However, we anticipate that the insulation effect is still sufficient for the dark matter to retain some of its heat to the present day, in a manner that would be possible for us to detect and then differentiate from other dark matter models.

\section{Field work}

We undertook two field trips to the eruption site in the spring of 2021, first on the 29th of March and the second on the 17th of April. The goals were to study the fresh lava as a WDM candidate, to take lots of photos, and to not fall off the side of any mountains on the trail / not be overwhelmed by gaseous volcanic fumes / not be struck on the head by falling rock that is both very heavy {\it and} has a temperature $>600$~K. 

\begin{figure}
	\includegraphics[scale=0.065]{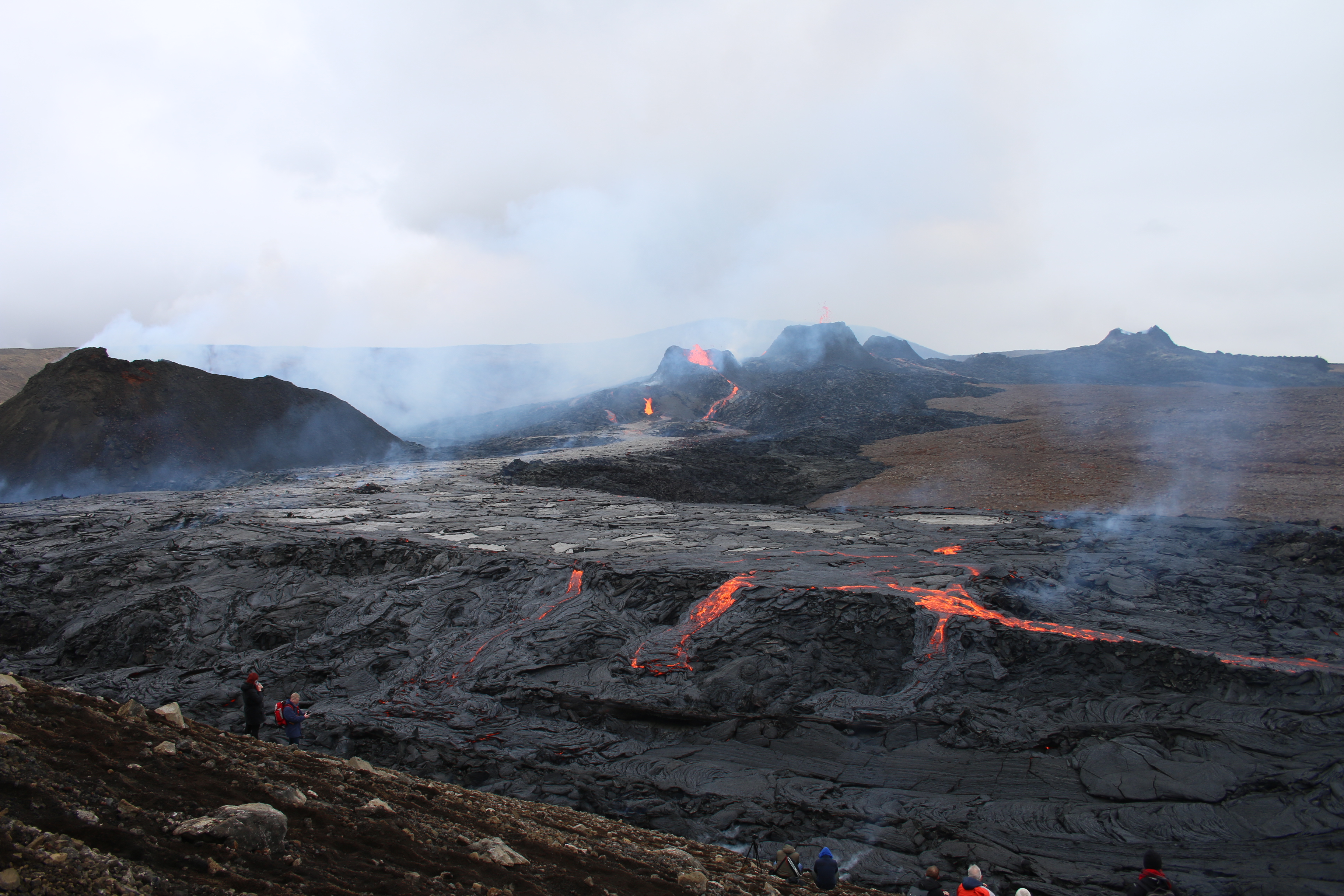}
	 \caption{The volcanic system on the 17th of April 2021. Note the smoothness of the top of the flow where small perturbations have been erased by the thermal motions of the lava.}
	 \label{fig:idd}
 \end{figure}

Having survived both legs of both trips, we can confirm that frozen lava is grey to black in colour, and thus a dark matter candidate\footnote{The word `dark' implies the absorption of light. Dark matter particles as typically considered do not interact with light at all, including absorption, so should arguably be more appropriately described as `transparent matter'. Lava is anything but transparent.}. We also verified that it is specifically a  WDM candidate, first from the smell of barbecue, second from the burning grass and third from the glow of lava immediately post-production and prior to any cooling.  We stated in the theory section that any WDM candidate should erase small perturbations; we present evidence of this effect in Fig.~\ref{fig:idd}, which is a photo taken during the second field trip.

The image features a crater row, and five of the eight craters generated over the course of the eruption are visible. Production of fresh lava occurs primarily in the two craters at the image centre, and subsequently flows towards the viewing location and on to the right-hand side of the image (an easterly direction). Crucially, we verify our hypothesis that lava is a WDM candidate that erases small scale perturbations. The remarkable flatness of the top of the flow does indeed indicate that the combination of lava viscosity and temperature is sufficient to remove perturbations smaller than the flow at large. The viscosity of the lava is set in part by its composition, and increases with the fraction of silicon in the lava. We therefore anticipate that future WDM studies will constrain the silicon fraction of dark matter, with more granular structure at small scales requiring larger amounts of silicon. Perhaps...
 
\section{Post field-trip observations}

The time interval between our last field work visit -- the 17th of April -- and the end of the eruption -- the 19th of September -- was characterised by large fluctuations in the lava production rate. Fortunately, the eruption was visible across the bay from a location just 30 seconds' walk from our front door, despite being 35~km distant. We were therefore able to monitor the progress of the lava emission and its suitability as a dark matter candidate with relative ease, provided the weather conditions were good. In this section we provide a summary of this behaviour.

\begin{figure}
	\includegraphics[scale=0.065]{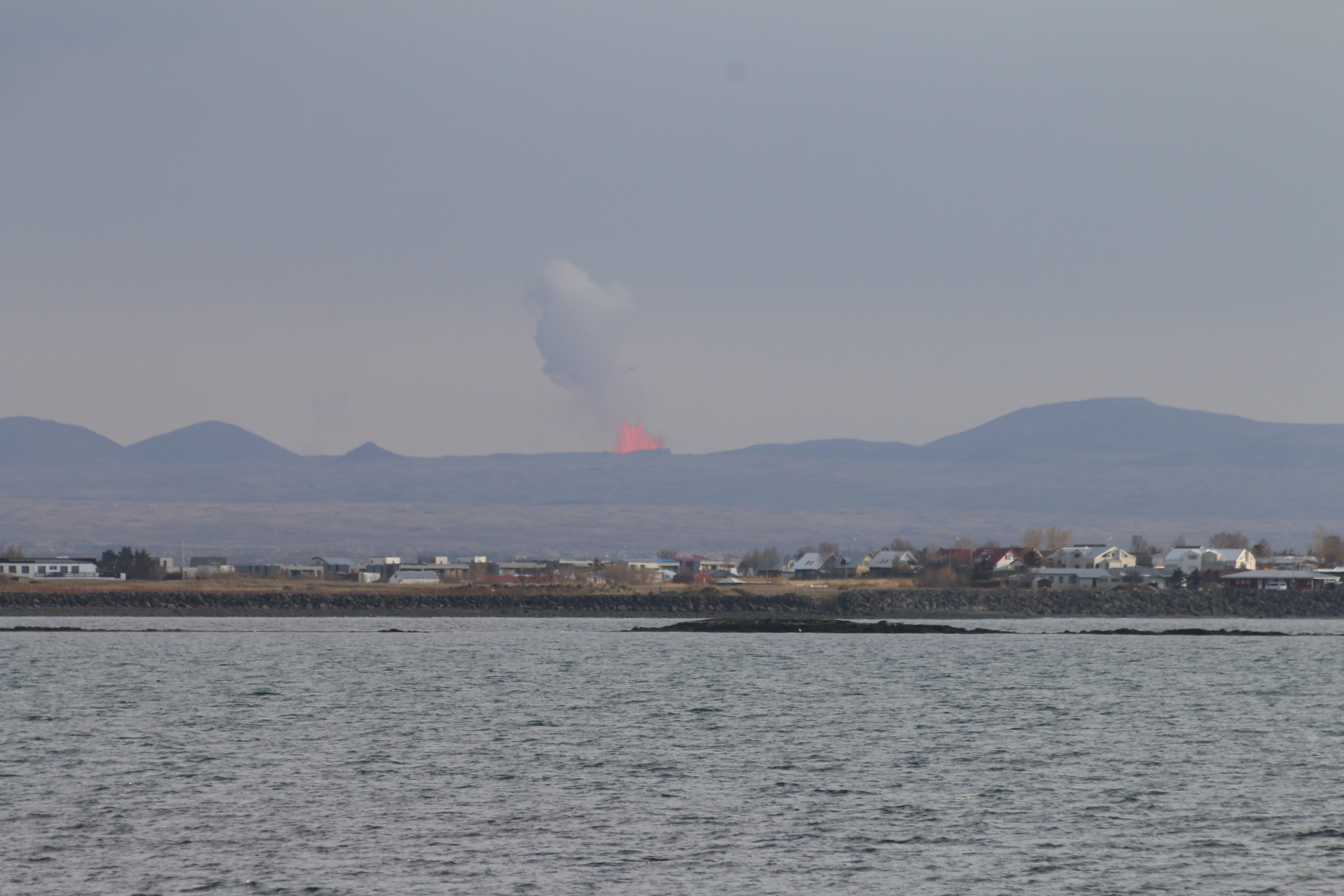}
	 \caption{Image of the volcanic crater and its surroundings on the 18th of May.}
	 \label{fig:lf1}
 \end{figure}

During May of 2021, the central crater from Fig.~\ref{fig:idd} grew to engulf most of the other craters and became the sole site of lava production. This crater then began a periodic eruption pattern with a duty cycle of 10 minutes, featuring 3 minutes of dramatic lava fountaining followed by 7 minutes of quiescence. An image of this behaviour is presented in Fig.~\ref{fig:lf1}, taken on the 18th of May. The height of the crater has grown to $\sim100$~m from its starting base in just two months, and the lava fountain reaches a further 50-100~m into the air. This would suggest that the production of lava WDM is sporadic; the consequences for galaxy formation are unclear. 

\begin{figure}
	\includegraphics[scale=0.065]{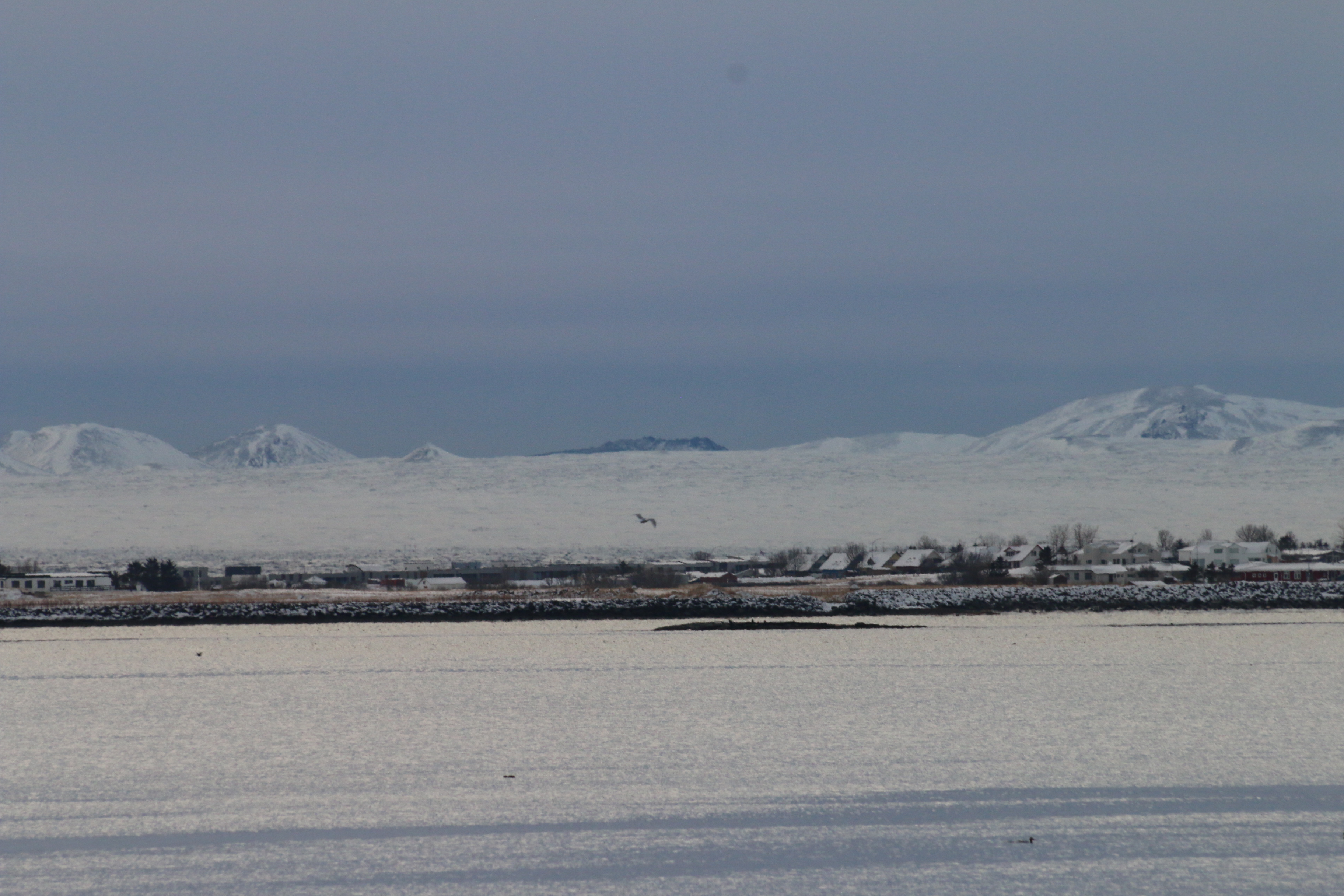}
	 \caption{Image of the volcanic crater and its surroundings on the 1st of December, approximately 1.5~months after the last lava flows were emitted.}
	 \label{fig:lf2}
 \end{figure}

A key feature of dark matter science is to infer the behaviour at early times from present day observations. CDM will show evidence that the dark matter has low thermal velocities at early times, whereas WDM will instead preserve its higher temperature. The last production event of lava was the 19th of September, thus in the subsequent months we had the opportunity to monitor post-production clues for lava as a dark matter candidate. One of the particular benefits of conducting this research in Iceland is the common occurrence of snowfalls from November to March, which can be used to measure dark matter temperature. This meteorological opportunity presented itself to us on the 1st of December, an image from which we present in Fig.~\ref{fig:lf2}.

The striking result of this image is that, while the surrounding mountains are covered in snow, the crater retains its dark colouring from the retention of its post-production heat. This finding corroborates our expectation from the theory section that this fresh lava is indeed a WDM candidate, and its warm nature is readily apparent months after production has ceased.

\section{Summary \& conclusions}
 \label{conc}
 
In this paper we considered the possibility that fresh lava is a WDM candidate, taking advantage of the 2021 volcanic eruption at Fagradalsfjall, Iceland to examine this hypothesis. To this end, we obtained evidence concerned with this hypothesis through two distinct channels: indirect observations at a distance of 35~km over several months and  a pair of field trips.  We demonstrated that fresh lava is indeed an ideal WDM candidate: the combination of 1000~K temperature and self-shielding from cooled flow outskirts enables the lava to exhibit free-streaming out of small perturbations. The self-shielding also enables the lava to retain its heat at least two months after the cessation of the eruption, which therefore leaves present days clues to lava's nature as WDM. 

 We also recognised that the ability of lava to flow is tied to its concentration of silicon, with silicon-rich lava more viscous, more resistant to free-streaming, and thus also trending more towards being a CDM candidate in some ways. Whether observations of galaxies can be used to constrain the silicon concentration of dark matter is a topic left to future work. 
 
\section*{Acknowledgements}

 MRL would once again like to thank the late Terry Pratchett for his contributions to the writing style applied in this paper. I would also like to thank Mother Nature for opening up a massive gash in the earth at a comfortable 20 miles from my apartment.                 

\bibliographystyle{mnras}
\bibliography{bibtex}
    
\bsp
\label{lastpage}

\end{document}